# Altermagnetizing the FeSe-like two-dimensional materials and approaching to giant tunneling magnetoresistance with Janus Cr$_4$BN(B$_2$) MBene electrode

*Weiwei Sun\*, Mingzhuang Wang\*, Baisheng Sa, Shaui Dong, Carmine Autieri\*, Zhongfang Chen*

Weiwei Sun, Shuai Dong

Key Laboratory of Quantum Materials and Devices of Ministry of Education, School of Physics, Southeast University, Nanjing, 211189, China. Email: W. Sun, provels8467@gmail.com

Weiwei Sun, Mingzhuang Wang

SEU-FEI Nano-Pico Center, Key Laboratory of MEMS of Ministry of Education, Southeast University, Nanjing, 210096 China. Email: M. Wang, 220226184@seu.edu.cn

Baisheng Sa

Multiscale Computational Materials Facility & Materials Genome Institute, School of Materials Science and Engineering, Fuzhou University, Fuzhou 350108, China.

C. Autieri

International Research Centre Magtop, Institute of Physics, Polish Academy of Sciences, Aleja Lotników 32/46, PL-02668 Warsaw, Poland. Email: C. Autieri, carmine.autieri@MagTop.ifpan.edu.pl

Zhongfang Chen

Department of Chemistry, University of Puerto Rico, Rio Piedras Campus, San Juan, PR, USA

**Abstract**: Altermagnetism is an emerging series of unconventional magnetic materials characterized by time-reversal symmetry breaking and spin-split bands in the momentum space with zero net magnetization. Metallic altermagnets offer unique advantages for exploring applications in spintronics as conductive metals allows for serving as electrode in magnetic tunneling junction (MTJ) and/or manipulation of spincurrent through external field. Through density functional theory calculations, the 2D altermagnet Janus Cr$_4$BN(B$_2$) was predicted. Both intra- and inter-layered magnetic exchanges between Cr atoms are in antiferromagnetic with the first four neighbours. The mixed ionic between Cr-B and covalent bond between B-N are formed in the layered material, which also exhibits an over 9 eV built-in work function differences. Leveraging the anisotropic spin-splittings with momentum dependency, like the FM magnet, we therefore designed $Cr_4B_3N$/vacuum/$Cr_4B_3N$ in-plane MTJs with a 7 Å thick barrie. We found that the Cr-B vertical edge-assembled electrodes based MTJs exhibited giant tunneling magnetoresistance (TMR) ratios of 91001%, by aligning the conduction channels of the electrodes in parallel and anti-parallel states in the momentum space. The proposed strategy of integrating the interlayer coupling as well as the N substitution to the FeSe-like $Cr_2B_2$ bilayer can also be mapped to other FeSe-like lattice for innovating a large spectrum of desirable magnets. Our work deepens and generalizes understanding toward altermagnetic 2D metallic electrode with a newly established metal boride (MBene), and broadens applications of the nanoscale spintronics.

"research article"

# 1. Introduction

Magnetic materials and their associated technologies have been instrumental in driving technological advancements and economic development over the past centuries, profoundly impacting various facets of human society.[1–5] Historically, magnetic compasses, for instance, played a crucial role in the age of exploration, enhancing maritime safety and trade efficiency and thereby driving global economic and cultural exchange. In the information age, the development of magnetic storage technologies has revolutionized personal and corporate data management while also providing essential infrastructures. One of the most prototypical and widely used spintronic device is the magnetic tunnel junction (MTJ), and in Julliere's theory, electrons maintain spin states one ferromagnetic (FM) electrode while quantum-mechanically tunnels into another FM electrode through an insulating barrier layer, which in turn results the tunneling magnetoresistance (TMR) effect.[6] The TMR effect is the core of MTJ, allowing for magnetic random-access memories (MRAMs), magnetic sensors, and other spintronic devices.[7–10] Therefore, a giant TMR ratio is imperative so as to achieve reliable reading and writing for the purpose of applications.

A new class of magnetic materials, known as altermagnets (AMs), have emerged as a promising alternative to FMs and conventional AFMs.[11,12] Different from conventional AFMs, certain magnetic space groups (MSGs) with the violated time-reversal ($\mathcal{T}$) symmetry allows AMs to exhibit alternating spin-split band structure in the momentum space, enabling these altermagnetic materials to behave similarly to ferromagnetic materials in the properties involving the k-space with locally spin-dependent channels defined as $p_{\parallel}(\vec{k}_{\parallel}) = \frac{D_{\parallel}^{\uparrow} - D_{\parallel}^{\downarrow}}{D_{\parallel}^{\uparrow} + D_{\parallel}^{\downarrow}}$. [12–21]

To date, many altermagnets have been discovered, but they are mostly found in the semiconducting state. [11] For example, $Fe_2Se_2O$ and $V_2SeTeO$ were predicted using density functional theory calculations, and both exhibited the multipiezo effect combining piezoelectricity with piezo-altermagnetism[22,23]. In ultrathin film of Weyl altermagnet GdAlSi, a spontaneous anomalous Hall effect manifests itself, indicating a nonrelativistic spin splitting.[24] Such altermagnetic effects also appear in semiconductors like $RuF_4$ and FeS(110).[25,26] Enriching more 2D AM magnets with distinct structures through manipulating the symmetry and chemical environment will not only deepen our understanding of unconventional magnetism but benefit to uncover new physics and principles of materials design.

AMs combining advantages of both FM and AFM exhibit peculia properties against magnetic perturbations and high-speed spin dynamics, making them ideal candidates for next-generation spintronic applications. [27,28] When a given AM is used as the electode in MTJs, the match of the momentum-dependent spin polarized channel is possible to achieve and induce outstanding conductance and TMR.[16] Recent studies have demonstrated that MTJs incorporating AM electrodes exhibit considerably high TMR, which is a critical metric for evaluating MTJ effectiveness. [16,29-34] In specific, TMR of ~500%, ~12000%, ~2%, ~6100%, 100%-200%, 299% have been theoretically reported in $RuO_2/TiO_2/RuO_2$ (001), $RuO_2/TiO_2/RuO_2$ (110), $RuO_2$ (110)/$TiO_2$/$CrO_2$, $Mn_3Se/MgO/Mn_3Se$ (experimental), $Ru_{1-x}Cr_xO_2/TiO_2/Ru_{1-x}Cr_xO_2$, $Mn_3Pt$/vacuum/$Mn_3Pt$, respectively.[16,29,31,33,35,36] Mapping a stable AM 2D layer to MTJ device to realize a giant TMR would be of significance to the revolution of the next-generation nano-spintronic devices.

Combining superconductivity with altermagnet is anticipated to make breakthroughs in the practical application of 2D magnetic materials. In fact, for the super-star of superconductor, the stripe (or double stripe) order in Fe(Se,Te) itself, [37] is not AM since the spin-mapping operation is a lattice translation as well as the spatial inversion. Interestingly, it can be tuned to the AM order with the manipulation of the electrical field, associated with quantized spin Hall (QSH) conductivity,[34] The similar observation also occurs in CaMnSi.[38] Apparently, the spontaneous AM state realized by the electrical field is not sustainable and only having slight spin-splitting in the momentum space. Meanwhile, the 2D Fe(Se,Te) lattice can be largely expanded to many other compositional combinations, and one of the derivative classes is the 2D metal borides, so-called anti-MXenes or MBenes in this report, more understandable in the perspective of chemical composition. [39] Moreover, the 2D metal borides can be stablized in a great number of lattices, e.g. orthorhombic, hexagonal, tetragonal, and trigonal lattices.[40] We also proposed several FeSe-like MBenes, like FeB, CoB, and IrB, which show interesting magnetic behavior and promising spin filtering effects.[41] Notably, the Cr-based MBenes is the mostly studied candidate for its stability and tunable magnetism. Therefore, proposing a feasible strategy to break the symmetry on this particular lattice could benefit the atomic design of 2D AMs with the same lattice and boost other interesting inherent phenomena like superconducting and quantum hall effects, which would significantly furnish the community.

In this report, based on the buckled FeSe-like bilayer (BL), we predicted a new 2D $Cr_4BN(B_2)$ MBene layer featuring janus two-sided surfaces to be the AM state. The transitions from the monolayer (ML), the N-substituted ML and the prototype BL and the optimzied structure are systematically overviewed and associated with the electronic structure. The charge transfer and bonding analysis between layers and the interatomic magnetic exchange couplings between Cr atoms within the first four nearest neighbors were thoroughly investigated. Non-Equilibrium Green's Function (NEGF) was employed to model the quantum transport properties of AFMTJs based on combinations of diverse vertical edge-assembled $Cr_4B_3N$ electrodes. We aim to advance the understanding of altermagnetic materials and their applications as electrodes in MTJs. These findings could pave the way for the development of next-generation MTJs having superior performance, thereby contributing to the broader field of energy-efficient and high-performance spintronics.

## 2. Results and discussions

### 2.1. Atomic structure and the dynamical stability

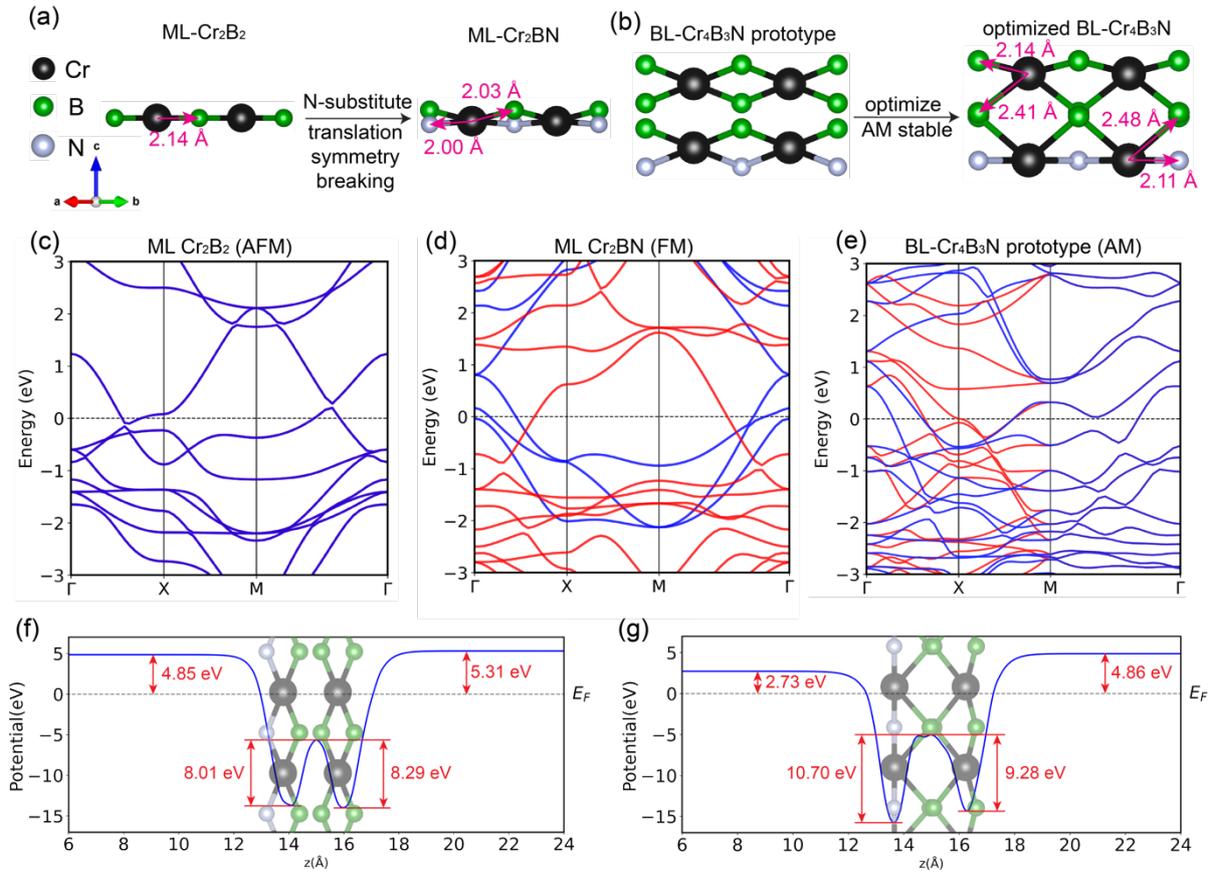

**Figure 1.** (a) Origin of translation symmetry breaking in ML-$Cr_2B_2$ and ML-$Cr_2BN$. (b) Optimized altermagnetic BL-$Cr_4B_3N$ obtained from the prototype constructed by buckled MXene layers, employing a combination of N substitution and bilayer stacking strategy. The green, black, and silver spheres represent B, Cr, and N atoms, respectively. (c-e) Band structure of (c) ML-$Cr_2B_2$ with conventional AFM ground state, (d) ML-$Cr_2BN$ with FM ground state, (e) BL-$Cr_4B_3N$ prototype with AM configuration. (f)-(g) The calculated work functions of the prototype and optimized BL-$Cr_4B_3N$.

The monolayer of CrB as a quasi planar,[41] is a conventional AFM ground state, as illustrated in the left panel of **Figure 1**a. For the ML-CrB, although AFM magnetic configuration has a lower energy, the P4'/n'm'm (#129.416) MSG does not violate the $\mathcal{T}$-symmetries. Therefore, we tend to lower the symmetry by breaking the inversion but preserving the rotational symmetry $C_{4z}$ and the mirror symmetry with respect to both the x- and y-axes. To do so, here, we utilize the N-substitution strategy to explore the potential existence of AM. However, the ground state of ML-Cr$_2$BN was calculated to be FM with a 0.02 eV/f.u. energetic favorability. The evolution of the bond length of Cr-B was tracked and showed that the involvement of N would shorten the bond with Cr (from 2.14 Å to about 2.0 Å), due to one more electron included. Moreover, the structure associated band structure is presented below the structural models.

In order to ground to an AM state while preserving the broken symmetry, we thus go beyond to the bilayer with one B substituted with N on one side, leaving the structure as shown on the left of Figure 1b.[42,43] Here, the inner surface boron layers do not tend to stay in a way of Van der Waals (VdW) but redistribute to form a quasi-plane of the octahedra with surface Cr atoms, as shown in **Figure 1**b. Moreover, the surface B-Cr-B with in 160° transforms to a N-Cr-N plane. We therefore claim that the interlayer interaction along with the N substitution lead towards the presented janus Cr$_4$B$_3$N. The calculated bands of the prototype Cr$_4$B$_3$N show the spin splitting (particularly at X), although rather weak, can be found in **Figure 1**e, which confirms our successful symmetry manipulation towards altermagnet. From the prototype to the optimized structure, we took into account the chromium positions and magnetic arrangements in this peculia structure, we setup our different magnetic configurations, AFM1, AFM2, AFM3 and FM in **Figure 2**a, from the left. The most stable magnetic structure can be determined by comparing their total energies. The energy differences between the confiugurations, i.e. $E_{AFM2}-E_{AFM1} = 0.10$ eV/f.u., $E_{AFM3}-E_{AFM1} = 0.26$ eV/f.u., $E_{FM}-E_{AFM1} = 0.83$ eV/f.u., presented in Figure 2b show that the most favourable magnetic confifuration is AFM1. Such thermodynamical differences are remarkably sizable, confirming the proposed AFM1 is quite stable and in strong competent against other magnetic configurations.

Before moving towards the band structure, the work functions of the designed prototype and optimized BL-Cr$_4$B$_3$N structures are computed and shown in Figure 1f-g. Due to the N-substitution, the two layers of the bilayer material exhibit distinct work functions, analogous to many other Janus 2D layers, like MoSSe and WSSe.[44,45] Calculations incorporating dipole corrections reveal that the two sides of the prototype has work functions of 4.85 eV and 5.31 eV, respectively. In contrast, the optimized configuration demonstrates work functions of 2.73 eV and 4.86 eV. Notably, the layer subjected to N-substitution exhibits a significantly lower work function compared to the other side. In a comparison with other MBenes with work functions generally between 4.1 and 5.3 eV, this newly established janus $Cr_4B_3N$ may render more conducive to electron emission.[41,46] For the layer resolved work fucntion, the extremely large built-in difference of the work function over 9 eV for both sides may stimulate more studies on the electrocatalyst by using of the reported janus MBenes.

Further more, the negative free PDCs shown in Figure 2c implies the lattice stability of the proposed structure with the AFM1 configuration. Combinng with the PHDOS shown on the right of Figure 1c, up to 200 $cm^{-1}$, Cr atoms leading phonons are dominant, because of its heavier mass than other elements inside the MBene. While the boron leading phonons are mainly distributed between 200 and 400 $cm^{-1}$, mixed with some Cr based phonon states, showing some dispersions. Beyond 400 $cm^{-1}$, the nitrogen leading phonons are emerging with localized or even quasi-flat charateristic. Note that two band gaps are placed to obscure the phonon transport, and the acoustic and optical coupling is quite evident, suggesting that the lattice thermal conductivity might be considerably supressed. Expanded study on the thermal conductivity may be plausible for investigating their thermoelectric applications.

## 2.2. Elctronic structure and magnetic exchange coupling

According to the symmetry analysis, the strcuture falls into a magnetic space group (MSG) Pm'm2' (#25.59), violating $\mathcal{T}$-symmetries.[47] The interlayer distances also differ due to the inequivalent chemical environment for the upper (2.41 Å) and lower layers(2.48 Å). Unlike conventional antiferromagnets, the sub Cr-lattices with opposite spin polarization are coupled by translation and/or inversion symmetry operations, the sub-lattices in bilayer $Cr_4B_3N$ are connected by the spin group symmetry$[C_2\|C_{4z}]$, as shown in Figure 2d, which breaks the time-reversal symmetry. $C_2$ represents the spin-space operation to flip the spin up/dn, and $C_{4z}$ represents a four-fold rotation along the z axis in real space. Such a symmetry is therefore satisfied with the alternating spin polarization in the momentum space.[47] The Fermi surface (FS) determines the disctribution of conductive electrons in the momentum space, thus playing provital role to the TMR ratio in MTJ as the electrode. The spin-dependent FS in Figure 2e. Owing to (4z|0,0,0) rotational operation, the spin-dependent bands and spin-up/down FS sheets also exhibits rotational symmetry. Two spin-up bands and one spin-down band cross the $E_f$ along the Γ-Y direction. While along the Y-M direction, two spin-down bands and one spin-up band cross $E_f$. Additionally, three spin-degenerate bands cross $E_f$ along Γ-M that are reflected in the FS sheets as well. Vividly, the FS can be dipcted as repeat bowlin-like sheets with pockets addressed in the interstitial regime along the path of Y-M.

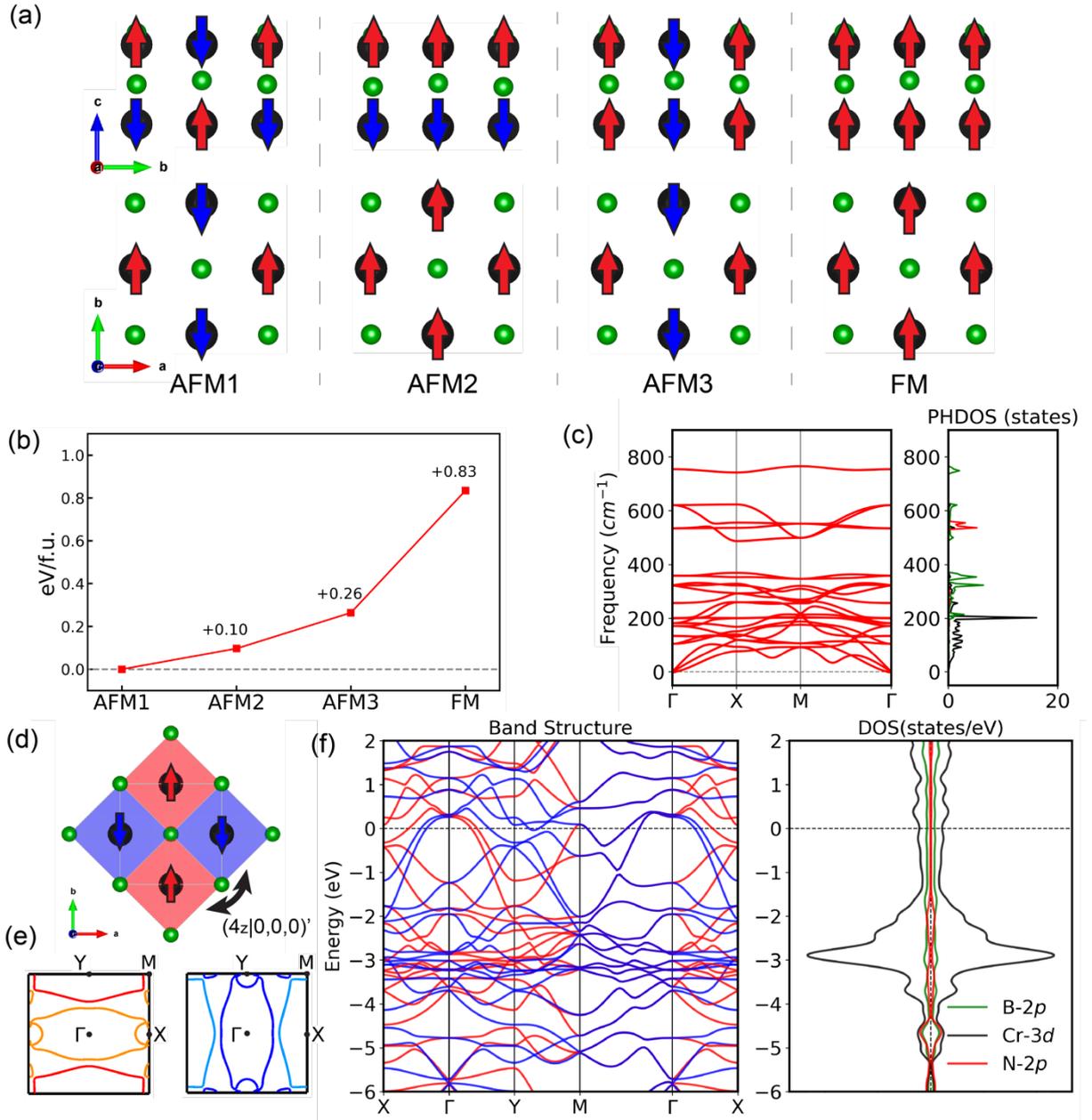

**Figure 2.** (a) Four different magnetic ground states considered for $Cr_4B_3N$. The red and blue vectors indicate spin-up and spin-down directions, respectively. Both side view (upper) and top view (lower) are diaplayed. (b) Relative energies of the four magnetic configurations of $Cr_4B_3N$. (c) The phonon dispersion curves (PDCs) along the path of $\Gamma$-X-M-$\Gamma$ and the asssociated phonon density of states (PHDOS). (d) The rotation symmetry operation in bilayer $Cr_4B_3N$. (e) The spin-up (left) and -dn (right) dependent Fermi surface, in which the color corresponds to the band index. The high symmetry points in the Brillouin zone are marked. (f) The spin-resolved band structure and projected density of states of B-2p, Cr-3d and N-2p states in $Cr_4B_3N$ are plooted.

Indeed, as shown in Figure 2f, the band structure of 2D $Cr_4B_3N$ exhibits the metallic characterisitic. The band dispersions long the X-$\Gamma$ and $\Gamma$-Y paths are identical but exhibit opposite spin spliiting that is clear and sizable. While energy bands are spin degenerat along the $\Gamma$-M in contrast to the bands along other *k*-paths, whose spin-up and -dn bands are split. The momentum-dependent spin-splitting with the above features reveal that the altermagnetism in $Cr_4B_3N$ has a d-wave spin texture. Moreover, the combination of peculiar *k*-dependent split and degenerate bands refer to the relianace of magnetic properties on the orientation, where direction-sensitive or controllable magnetic properties can be harnessed for magnetic sensors and high-density data storage.

The QSH topological phase in Fe(Se,Te) type of structure, the parental materials of 2D $Cr_4B_3N$ is protected by the presence of mirror reflections with respect to both the x- and y-axes[48], and the same mirror symmetries can also protect QSH topological phases in multilayer systems. Consequently, multilayers can host the necessary symmetries for realizing the QSH conductivity, provided that the band inversion occurs at the Fermi energy ($E_f$) and no trivial bands are present at $E_f$. However, the large bandwidth in the MBene poses a challenge to the straightforward realization of the QSH phase. Therefore, the topological properties will not be further explored in the present report.

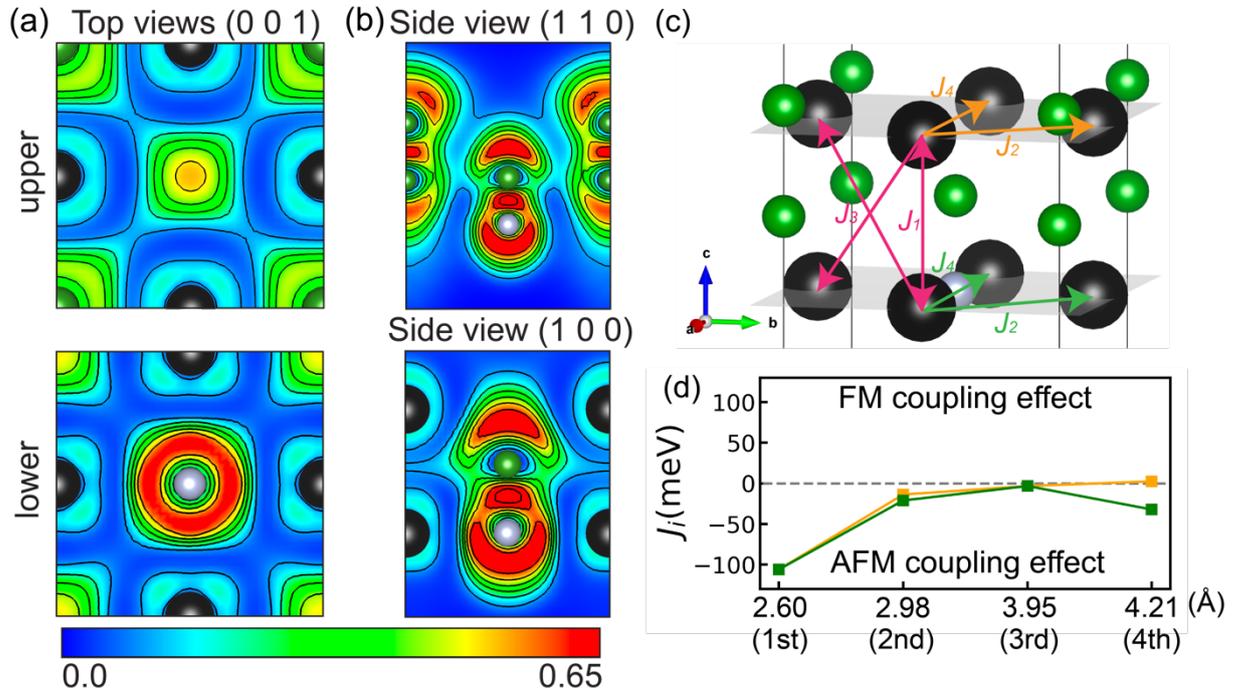

**Figure 3.** (a) The electron localization functions (ELFs) projected into (001) planes. Two diagrams are corresponding to the uppder and lower Cr layers, respectively. (b) ELFs projected into (110) and (100) planes. The distances from the origin of these two planes are is 1 $\vec{a}$ and 0.5

$\vec{a}$. (c) The schematic model of the exchange interaction between neighboring Cr atoms. $J_1$ to $J_4$ represent the nearest (1st) to forth nearest neighbor (4th) magnetic exchange coupling. The orange and green arrow and labels denotes the exchange interaction for atoms in the upper and the lower layer, respectively. (d) The calculated magnetic exchange parameters as a function of nearest neighbors. The black, green and white atoms represent Cr, B and N, respectively. The presented ELF projection is within the range bwteeen 0-0.65.

Enlightened from the structure of $Cr_2B+B_2+Cr_2N$ along the z-axis, the distinct coordination of Cr-B and Cr-N in the two surfaces would lead to a significantly occupational differences of in the $d_{x^2}$ orbital, compared to other orbitals. Actually, in our DFT simulations, the two types of Cr atoms are with local moments of 3.7 μB and 3.5 μB, showing rather small differences. For the spin-up channel, $d_{xy}$, $d_{yz}$ and $d_{z^2}$ orbitals of the Cr atom are approximately in equivalent occupancies of 0.8 $e^-$, and $d_{xz}$ and $d_{x^2}$ orbitals are approximately 0.7 and 0.6 $e^-$ occupied. In the other spin channel, merely the $d_{x^2}$ orbital is fractional occupied with 0.2 $e^-$. The anisotropic occupations in the five d orbitals are partially determined by its low symmety.

It is puzzling that the varying chemcial environment of the two types of Cr atoms are in rather similar local moments and occupations for the d orbital. Then, we may need the aid of the bonding charaterisitics to elucidate such an interesting phenomena. The intra- and inter-layer interactions in $Cr_4B_3N$ can be understood and interpreted by examining the electron localization functions (ELFs). Here, we mapped it to two (001) and another two (110) planes illustrated in **Figure 3**a. In both upper and lower layers, Cr atoms are in the ionic interactions with boron and nitrogen atoms, agreeing well with the observed DOS in **Figure 2**f, where no strong hybridization between Cr-3d and B-2p states. While the B atoms in the upper layer do form weak covalenct bond with the central B atom in the middle plane, indicated by the shared cloud in the cyan colour, as shown in **Figure 3**b. Combining the DOS plot shown in Figure 2c, the Cr-3d states are primarily located at -3 eV without strong hybridization with either B-2p or N-2p states, in conjunction with the pronounced DOS peaks in the range of [-4, -2] eV indicating a characteristic of ionic bond. As the Cr-3d states prevail to around -5 eV, evident hubrization between the Cr-3d and B/N-2p states are observed, suggesting that the boron and nitrogen atoms are covalently bonded and exhibiting relatively strong ELF in between on the right panel of Figure 3b. Indeed, the vertically aligned B-N has very localized ELF in between. The central B atom in the middle does have widespread ELF cloud outward, forming a weak covalent bond with the B atoms in the upper layer, visible in **Figure 3**a. This might recall us that in Figure 2c, the B-2p and N-2p states are also in strong overlapping. In summary, the sandwiched $Cr_4B_3N$ layer combines ionic and covalent bonding for the B-N bond, therefore characterized by a mixed bonding. Such anomalous covalency between anions are in strong contrast to the conventional M-X, deepening our understanding of the so-called anti-MXenes.[39]

Moreover, the formation of the transition-metal boride 2D layer are intricately linked to variations in different filling configurations of these d-bands in Cr, giving rise to unique magnetic behaviors of these MBenes. The electronic structure of metal borides is strongly influenced by the of the transition-metal atoms, i.e. the number of their d-electrons as well as coordination environment. At the beginning, assuming that all constituent elements of the MBene are in their nominal oxidation states ($B^{-5}$/ $B^{-1}$, $N^{-3}$), the Cr–B and Cr–N bonding states are both filled, while their antibonding states remain empty. Hence, only electrons occupying the nonbonding d-orbitals can be assumed to contribute to magnetism. For the structure of $Cr_4B_3N$, two types of boron atoms are found, two boron atoms in the middle plane (~2.4 Å for the Cr-B bond) and the other one situated in the top plan tightly binding with the two Cr atoms (2.1 Å for the Cr-B bond). We thus use -1 (the former) and -5 (the latter) oxidation states for the two different types of boron atoms. The averaged nominal oxidation state of each Cr ($[Ar]3d^54s^1$) in $Cr_4B_3N$ is counted to +2.5 after donating five electrons to one boron atom and one electron to boron atoms in the middle plane, and three electrons to one N in the center of the bottom plane. Following Hund's rule, the remaining three and half d-electrons fill the d band, which gives rise to a local magnetic moment of 3.5 μB per Cr atom. As discussed above, the DFT calculated local moments are agreeing well with the above analysis. It is worth noting that such simplified model also works for many other MXenes.[49]

To define the type and the strength of magnetic exchange coupling, we employ the Heisenberg model:

$$H = -\sum_{i\neq j} J_{ij}\hat{e}_i\hat{e}_j \qquad \text{Eq. 1}$$

where $\hat{e}_i$ is the normalized local spin vector on atom *i*, and $J_{ij}$ is the Heisenberg exchange coupling constant. This was achieved by a real-space Green's function implementation of the original Liechtedstein-Katsnelson-Antropov-Gubanov (LKAG) formula:[50–52]

$$J_{0i,Rj} = -\frac{1}{4\pi}\int_{-\infty}^{E_F} d\epsilon\, ImTr[\Delta_i G_{i,j}(\epsilon,\boldsymbol{R})\Delta_j G_{i,j}(\epsilon,-\boldsymbol{R})] \qquad \text{Eq. 2}$$

Where *i* and *j* label atomic indices within a unitcell, **R** is a lattice vector, and $\Delta_i = H_i^\uparrow - H_i^\downarrow$ is the on-site difference between the up and down part of the Hamiltonian matrix.

As shown in **Figure3**c-d, the interlayered distance between Cr-Cr atoms is measured to be the shortest one, followed by the inter- and intra-layer ones. Enlightened from the distance dependent $J_{ij}$, the exchanges between 1st to 4th neighbors are all in AFM coupling with negative values for the four $J_{ij}$, with an exception of $J_1$ for the upper layer. The $J_1$ coupling between the 1st nearest neighbors as of -105.9 meV ensures the antiparallel coupling between the upper and lower layers. And the exchange couplings between the 2nd nearest neighbors in the upper and lower layers exhibit rather closed, which are -13.7 meV for the former and -21.0 meV for the latter. This corresponds to a result of similar charge states of Cr atoms in the upper and lower layers, as illustrated in the Bader charge analysis above. Interestingly, couplings between 4th nearest Cr neighbors significantly differ. $J_4$ for the upper and lower layers are calculated to be 2.5 meV and -32.1 meV. This is primarily due to and the peculiar covalent bonding in B-N and the obscure N atom as visible in Figure 3a,b. In a nutshell, the distance dependent $|J_i|$ diminishes with going to longer range interactions, and the AFM magnetic exchanges are always predominant, illustrating that the alter-AFM state is strong and robust. The presence of the B-N colvlent bond makes the $J_3$ in the two layers significantly different, but $J_2$ in the two layers are rather closed due to the similar chemical environment and charge states.

### 2.3. Magnetic tunnel junctions and tunneling magnetoresistance

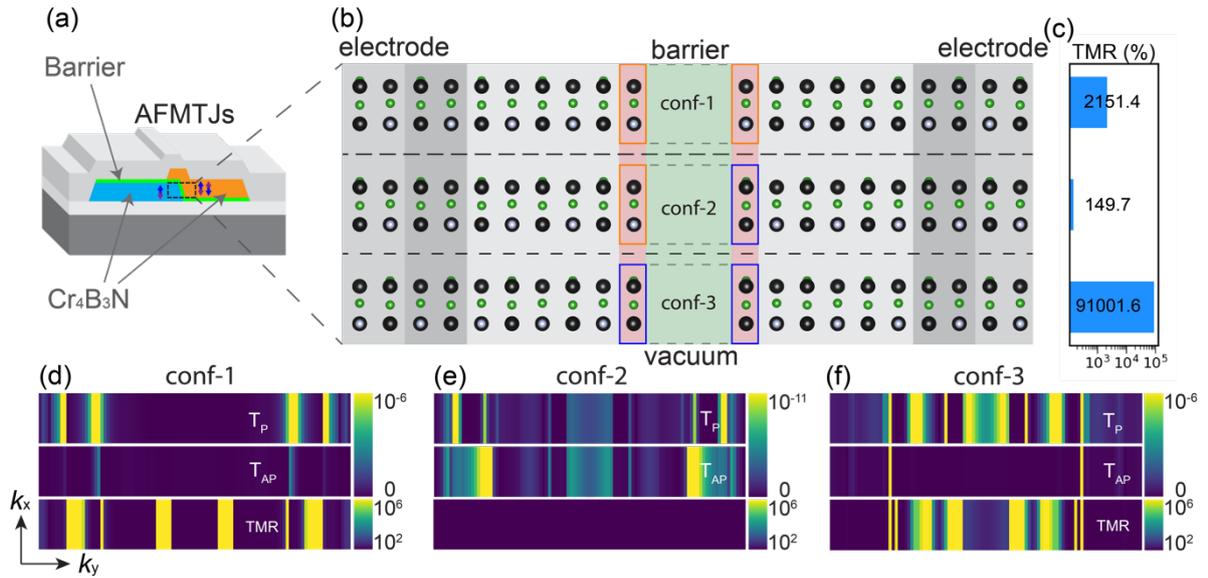

**Figure 4.** (a) Schematic representation of a MTJ device based on bilayer $Cr_4B_3N$ (b) Side view of atomic models of $Cr_4B_3N$/vacuum/$Cr_4B_3N$ MTJ for three different edges. The edges of Cr-B and Cr-B-N are marked in blue and orange, respectively. The conf-1 has Cr-B-N edges on both electrodes; conf-2 has Cr-B-N and Cr-B edges, respectively; and conf-3 has Cr-B edges on both electrodes (c) The calculated TMR value of three MTJ configurations. The $\vec{k}_{||}$-resolved transmission and TMR in the 2D Brillouin zone of $Cr_4B_3N$/vacuum/$Cr_4B_3N$ for (d) conf-1 (e) conf-2 (f) conf-3.
12

Due to its matallic properties and substantial spin-splitting in the momentum space, we have considered an in-plane MTJ device based on bilayer $Cr_4B_3N$, as illustrated in **Figure 4**a. Three types of $Cr_4B_3N$/vacuum/$Cr_4B_3N$ MTJs can thus be established and developed with varying edges, with symmetryic Cr-B (conf-3) and Cr-B-N edges (conf-1) on both sides as well as asymmetric edged electrodes (conf-2). Without the loss of generality, a vacuum barrier of constrained 7.0 Å was selected, as the atomic structures depicted in Figure 4b. Although both electrodes are made of material that doesn't have net magnetic moments like the FM ones, the momentum-dependent spin polarization is anticipated to generate a spin-polarized current, which could also lead to pronounced TMR effects, defined by $TMR = (T_P - T_{AP})/T_P$. Note that P and AP denote the parallel and antiparallel states.

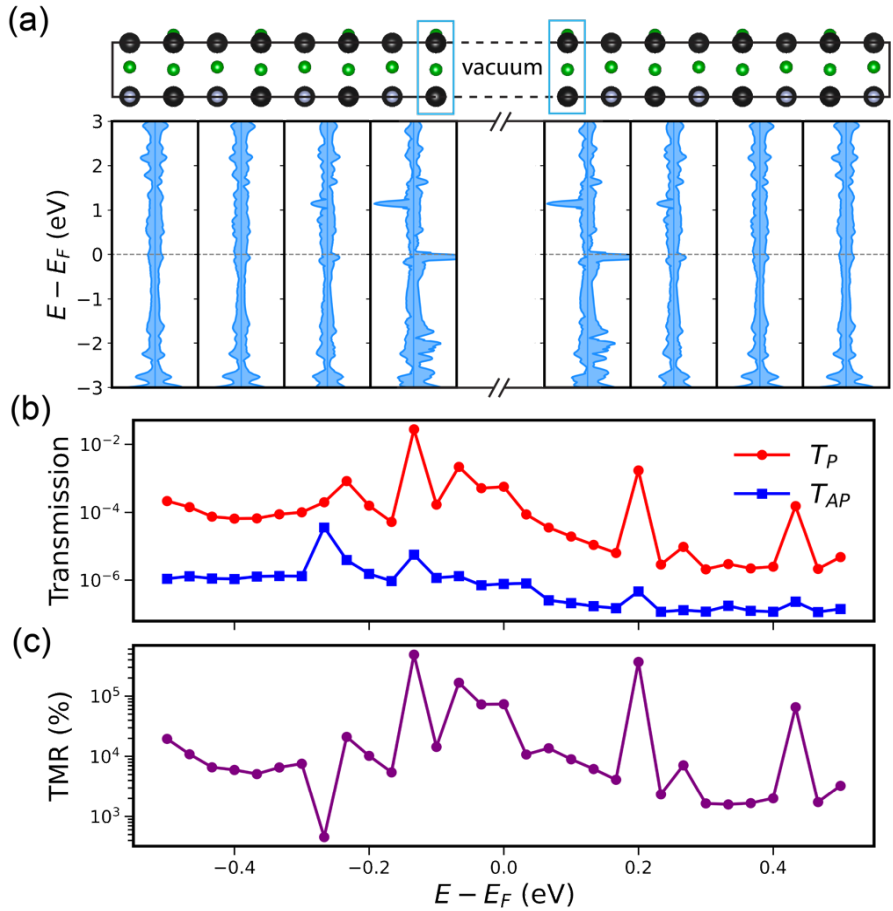

**Figure 5.** (a) The atomic model of conf-3 with the Cr-B edge on both electrodes and the corresponding bilayer-resolved density of states with the spin polarization. Each panel represents the DOS of a given bilayer, and panels from left to right correspond to the MTJ model. (b) Total transmission and TMR as a function of energy for the case of conf-3.

Transmission calculations are performed for the spintroic device application based on the newly established 2D layer, and the TMR ratios are displayed with each device configuration shown in Figure 4c. Among these configurations, conf-2 with the asymmetric edges with Cr-B-N on one side and Cr-B on the other side exhibits the smallest TMR. This may stem from the fact that the distribution of T$_P$ is almost the same as the transmission of the AP state. As a result, these two states have quasi-identical conductance patterns, leading to an almost zero TMR ratio of about 149%. While the conf-1 model with Cr-B-N edges on both electrods exhibits the modest TMR as approximately 2151%. Such a decent TMR ratio can be attributed to the uncompensated P state and the AP state, particularly the almost unconductive AP state in the entire presented **k$_x$-k$_y$** plane, leaving for a high conductance in the P state. Therefore, this ATMTJ with Cr-B-N edges based MTJ exhibited a sizable TMR effect at the level of 2000%, which is rather close to but slightly enhanced compared to Fe/MgO/Fe-based MTJ.[53] Notably, conf-3 with Cr-B edge on both sides could achieve a TMR as high as 91001.6%. Now, the current cannot flow through the **k** space of the AP state but induce several transmission strips in the P state, causing an ultra-high TMR, which is clearly higher than that of conf-1 by four orders of magnitude. Moreover, this ultra-high TMR is superior to all known AFMTJs reported so far, as the best performance lies in RuO$_2$/TiO$_2$/RuO$_2$ (110) as reported in Ref. [14], with a TMR of approximately 12000%. The ultra-giant TMR ratio in comparison to other AF- and FM- MTJs greatly meets the needs for the MRAM and sensor applications and shows great potential as the next-generation of nano-spintroic devices.

The $\vec{k}_{||}$-resolved transmission patterns for parallel $T_P^\sigma(\vec{k}_{||})$ and anti-parallel $T_{AP}^\sigma(\vec{k}_{||})$ states for the three configurations are presented in Figure 4d-f. Due to the 2D bilayer structure and the specific edge, $\vec{k}_{||}$-resolved transmission appears as strips, but we can distinctly observe the difference between the P and AP states. Moreover, the matched spin transmission of different edged electrodes can be sophisticatedly tuned. For the conf-2, the matched spin channels lead to the extremely weak transmission (in an order of $10^{-11}$) along with a larger κ↓ (the boundary regime), resulting in an exceptionally suppressed TMR. Such observation reflects the asymmetry in the distribution of the spin-polarized conduction channels in Cr$_4$B$_3$N. In other words, even with the varying edged electrodes, almost identical transmissions in the P and AP states can be obtained. While for the conf-1 and conf-3, a significant difference in transmission between P and AP states can be observed under $10^{-6}$ limit, and an enormous TMR exceeding $10^6$ % can be achieved. While the varying edges result in different tunneling regime. For the case of conf-1, the large transmissions occur merely at the **k** plane boundary in the AP state, and in conf-3, several discrete strips take place at the central regime along the **k$_y$** axis, exhibiting considerable transmission. The large transmission in both cases are seen in the P state. For the AP state, the transmission $T_{AP}^\sigma(\vec{k}_{||})$ is almost blocked for the wave vectors $\vec{k}_{||}$ with no conductance. Thus, thanks to the large difference of edge dependent $\vec{k}_{\parallel}$-resolved transmission between the P and AP states, we can generate ultra-giant TMR in conf-3.

Apparently, the transport properties of the newly established ATMTJ discussed above were calculated at the $E_f$. To further investigate the influence of doping deficiency or other effects that can induced $E_f$ shift (if any) on the TMR ratio, we extended the transmission and TMR ratio of 2D bilayer $Cr_4B_3N$/vacuum/$Cr_4B_3N$ AFMTJ of conf-3 to a energy window of [-0.5, 0.5] eV, shown in **Figure 5**a. For the bilayer dependent DOS, near the barrier, a high peak appears at the $E_f$, and some hybridiations in the range of [-3, -2] eV are present, and the DOS in the two spin channels are inequivalent, inducing net moment. To the electrode side, the large inequivalences in DOS of the two spin channels gradually vanish, as seen from the left to right pannles in Figure 5a. Furthermore, the almost identical transmission difference in the P and AP states with the ultra-giant TMR would lead to an excellent and stable performance for a nano-MTJ, as suggested by Figure 5b. Note that the TMR effect mainly generated between -0.2 eV to 0.2 eV as shown in Figure 5c, and the proposed MTJ could operate effectively across a wide voltage range sustainably between the P and AP states with a giant TMR.

Overall, the proposed AFMTJs with same edged electrods based on the 2D $Cr_4B_3N$ altermagnet would generate a remarkably stable and significant TMR effect, which is caused by the match of spin channel and also friendly for experimental fabrications and attempts. To achieve a stable and high-performance device, further research on edge control techniques and the selection of barrier layers is necessary, as with other AFMTJs.[31] The predicted TMR remains incredibly high and stable within the range of $E_f \pm 0.5$ eV, indicating that external factors may not cause evident impact on the operating functionality. In addition, the giant magnitude of TMR indicates a possibility of a strong spin transfer torque in the AFMTJs, which may offer an alternative way to switch the Néel vector.

## 3. Conclusion

In conclusion, we successfully employed DFT to predict a robust 2D AM magnet in a new janus MBene $Cr_4 B_3NB_2$. The discovered janus MBene is featured on two different surfaces, one in plain and the one in the zigzag quasi-plane, joint by octhedra backboned with the middle B layer. Substituting one central B atom with a foreign atom, N in this case breaks the translational symmetry, and the layer-layer interaction stablizes anti-ferromagnetic states both as a fundamentally key strategy to achieve the altermagnetism for a great number of FeSe-like layers. The AFM magnetic exchange is found to be predominant for both interlayer and intralayer interactions, showing that the AM state is quite robust and stable. The anomalous mixture of bonding, comprising both ionic and covalent characteristics, is demonstrated to uphold the agreement between the estimated and calculated local magnetic moment of about 3.5 $\mu$B. The large difference of the bulilt-in work function of more than 9 eV between surface and middle layers for both sides are identified and likely to make great contributions to the electrocatalyst.

We further investigated the tunneling process through the AM electrode with a vacuum as the barrier layer assembled $Cr_4B_3N$/vacuum/$Cr_4B_3N$ in-plane MTJ. The edge-dependent transmission in the momentum space in the P and AP states could considerably change the resulting TMR, with two candidate edges: Cr-B-N and Cr-B. It is worth noting that the MTJ with Cr-B edged electrodes on both sides reported here exhibits a giant TMR of 91001%, due to the almost zero The $\vec{k}_\parallel$-resolved transmission in the AP state. Furthermore, the predicted TMR ratio remains high and steady within the range of $E_f \pm 0.5$ eV, revealed by the transmission in both P and AP states. Thus, these MTJs with the same edged MBene showing the AM state have great potential for applications in the fields of MRAM and magnetic sensors. Such an excellent performance may also enhance its tolerance against to external stimulus, particularly helpful for devices in extreme conditions. Overall, this report provides a promising approach to exploring the electrical and transport properties of 2D AM electrodes while promoting the development and application of unconventional AM spintronics.

## 4. Methods

The first-principles calculations of the atomic and electronic structures are performed based on density functional theory (DFT)[54] as implemented in the Vienna *ab initio* simulation package (VASP) with post-processing using VASPKIT.[55,56] The pseudopotentials are described using the Projector Augmented Wave method, and the exchange-correlation functional is treated within the generalized gradient approximation (GGA) developed by Perdew-Burke-Ernzerhof (PBE).[57,58] In the calculations, the cutoff energy for the plane-wave expansion is set to 600 eV, and 29 × 29 × 1 Monkhorst-Pack grids k-point is set to sample the irreducible Brillouin zone.[59] The total energy is converged to $1 \times 10^{-6}$ eV atom$^{-1}$, and the error of force on each atom is less than $1 \times 10^{-2}$ eV Å$^{-1}$. For the 2D layers examined in this work, the lattice vector along the *z* direction was fixed to 30 Å to avoid spurious interactions between periodic images. Dudarev's approach was employed to evaluate the strong correlation effect on the d-orbital, with the correlation strength represented by the effective Hubbard $U_{eff} = U - J$ on the d-orbitals, where *U* and *J* are the on-site Coulomb and exchange parameters, respectively.[60,61] The $U_{eff}$ was set to 4.0 eV on Cr 3d-orbitals, which is estimated with reference to the literature on $Cr_2B_2$ XBene.[62] The phonon spectra calculations were carried out with the PHONOPY code in the framework of the density functional perturbation theory (DFPT) on a $3 \times 3 \times 1$ supercell with $5 \times 5 \times 1$ Monkhorst-Pack grids k-point to access the dynamical stability.[63,64] The magnetic space group was analyzed with the help of amcheck[65] and Bilbao Crystallographic Server.[66–68] The Fermi surface was calculated by QuantumATK, and visualized by FermiSurfer.[69] The exchange coupling parameters are calculated using a real-space Green's function implementation of the original Liechtedstein-Katsnelson-Antropov-Gubanov (LKAG) formula, as implemented in QuantumATK.[50–52]

Transport properties are calculated using the non-equilibrium Green's function formalism (DFT+NEGF approach)[70,71], as implemented in QuantumATK[72], using the atomic structures relaxed by VASP. In QuantumATK, the non-relativistic Fritz-Haber-Institute (FHI) pseudopotentials using a single-zeta-polarized basis were used in the device transport calculation. The spin-polarized GGA+U functional with $U_{eff}$ = 4.0 eV on Cr 3d-orbitals is included in the calculation. A cut-off energy of 200 Ry and a $1 \times 15 \times 225$ k-points density are used for the self-consistent calculations to eliminate the mismatch of Fermi level between the electrodes and central region. A $101 \times 101$ k-points grids for the $k_x$-$k_y$ space was used in the transmission calculation to reach a high accuracy.

**Supporting Information**

Supporting Information is available from the Wiley Online Library or from the author.

**Author Contributions**

**Weiwei Sun** and **Carmine Autieri**: Conceptualization, Formal analysis, Writing-review and editing, Funding acquisition. **Mingzhuang** Wang: Investigation, Visualization, Writing-original draft. **Shuai Dong**: Resources, Validation and Writing-review and editing. **Baisheng Sa**: Software, Methodology and Writing-review and editing. **Zhongfang Chen**: Validation, Methodology and Writing-review and editing. All authors read and approved the final version of the manuscript.

**Acknowledgements**

C.A. was supported by the Foundation for Polish Science project "MagTop" no. FENG.02.01-IP.05-0028/23 co-financed by the European Union from the funds of Priority 2 of the European Funds for a Smart Economy Program 2021–2027 (FENG). Part of the work was supported by the National Natural Science Foundation of China (No. 12234005 and T2321002). W.S. also acknowledges the Big Data Computing Center of Southeast University for providing the facility support on the numerical calculations in this paper.

**Conflict of Interest Statement**

The authors declare that they have no known competing financial interests or personal relationships that could have appeared to influence the work reported in this paper.

**Data Availability Statement**

The data that support the findings of this study are available from corresponding authors upon reasonable request.

**Ethical Statement**

Not applicable.
acknowledges the Big Data Computing Center of Southeast University for providing the facility support on the numerical calculations in this paper.

**Conflict of Interest Statement**

The authors declare that they have no known competing financial interests or personal relationships that could have appeared to influence the work reported in this paper.

**Data Availability Statement**

The data that support the findings of this study are available from corresponding authors upon reasonable request.

**Ethical Statement**

Not applicable.

Our work proposes a 2D altermagnetic MBene $Cr_4B_3N$ and provide an ultrahigh TMR ratio of 91001% with the capability of operting in a widely applied voltage of $E_f \pm 0.5$ eV, deepening the fundation of altermagnetic 2D material and laying the fundation of a realistic strategy to design high-performance magnetic tunnel junction.

Weiwei Sun*, Mingzhuang Wang*, Baisheng Sa, Shaui Dong, Carmine Autieri*, Zhonfang Chen

**Altermagnetizing the FeSe-like two-dimensional materials and approaching to giant tunneling magnetoresistance with Janus $Cr_4BN(B_2)$ MBene electrode**

ToC figure ((Please choose one size: 55 mm broad × 50 mm high **or** 110 mm broad × 20 mm high. Please do not use any other dimensions))

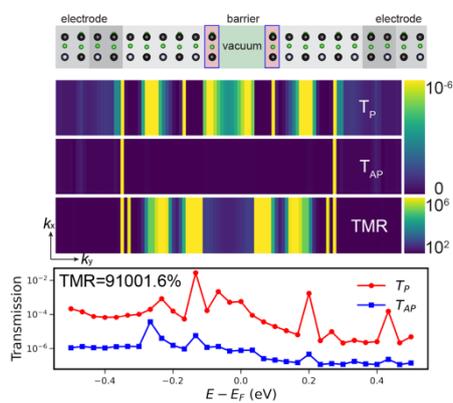

Supporting Information

**Altermagnetizing the FeSe-like two-dimensional materials and approaching to giant tunneling magnetoresistance with Janus $Cr_4BN(B_2)$ MBene electrode**

The geometry optimization nd symmetry analyses were both performed to monolayer $Cr_2B_2$ and $Cr_2BN$. Neither of the two exhibits AM behavior. In the case of $Cr_2B_2$, the energy of FM configuration is 0.31 eV/cell higher than that of AFM configuration, and for $Cr_2BN$, the energy of FM configuration is 0.02 eV/cell lower than that of AM configuration. For the monolayer $Cr_2B_2$, although AFM magnetic configuration has lower energy, the P4'/n'm'm (#129.416) MSG does not violate the $\mathcal{T}$-symmetries. The monolayer $Cr_2B_2$ can be ruled out as an AM candidate. As illustrated in **Figure S2**, it is a conventional AFM.

In the case of monolayer N doped CrB, the energy of AFM configuration of $Cr_2BN$ is energetically higher than that of the FM conterpart, indicating that it favours the FM state, as illustrated in **Figure S3**.

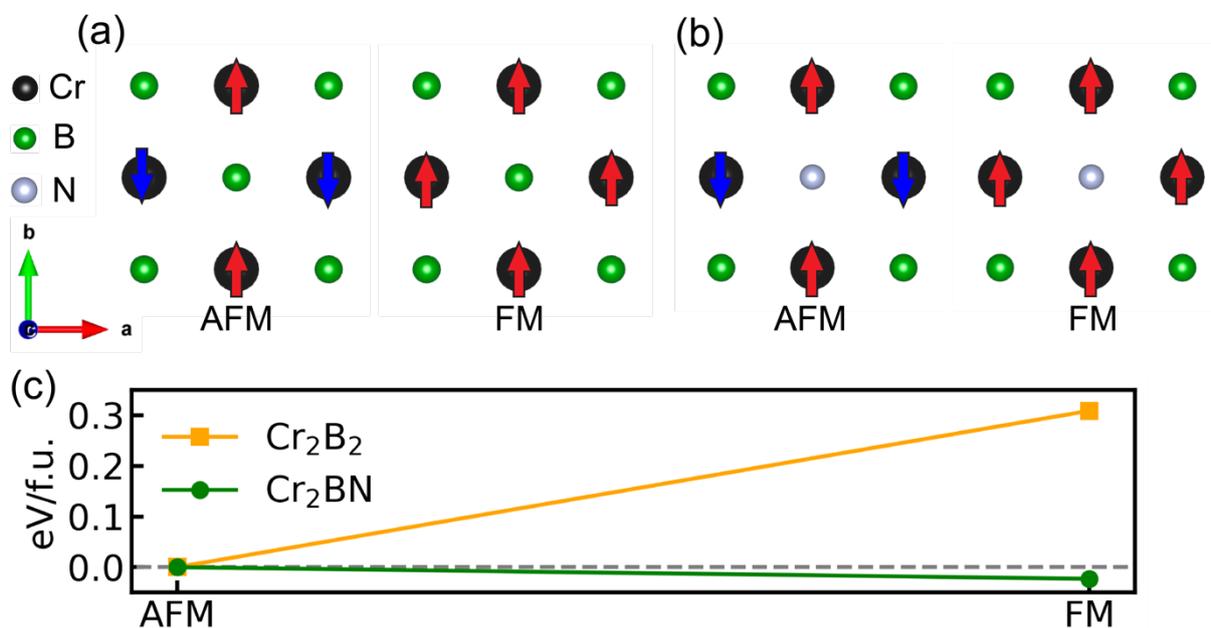

**Figure S1.** The top view and two magnetic configurations for (a) $Cr_2B_2$ (b) $Cr_2BN$, respectively. (c) The calculated energies of different configurations. For $Cr_2B_2$ the energy difference between the FM and AFM configurations is $\Delta_{FM-AFM} = 0.31$ eV/cell. For $Cr_2BN$, $\Delta_{FM-AFM} = -0.02$ eV/cell.

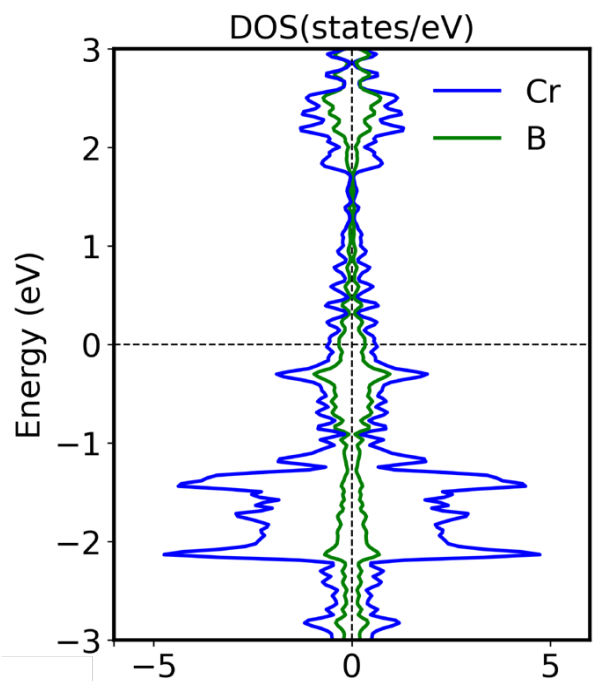

**Figure S2.** The density of states (DOS) of monolayer CrB, exhibiting conventional AFM properties.

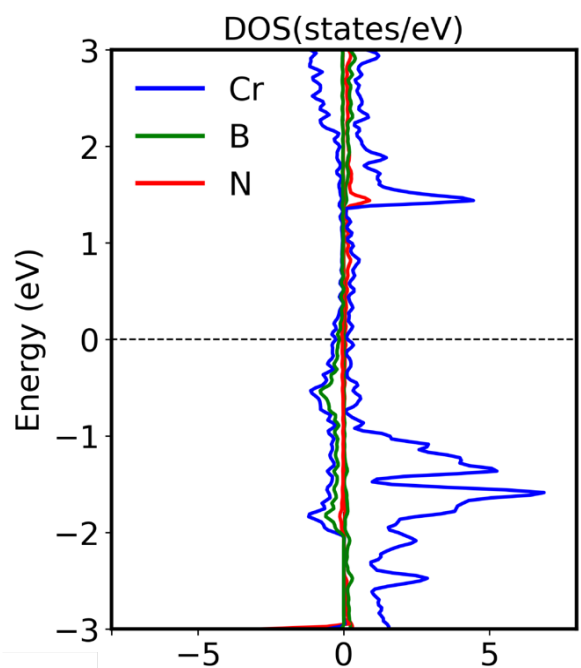

**Figure S3.** The density of states (DOS) of monolayer Cr$_2$BN, exhibiting FM properties.